\documentclass[letter]{aa}
\usepackage{graphicx}
\usepackage{txfonts}
\usepackage{natbib} 
\usepackage{psfrag}
\bibpunct{(}{)}{;}{a}{}{,}

\def\uJy{$\mu\mathrm{Jy}$}

\begin{document}
	\title{Rapid variability of the compact radio sources in Arp220}
	\subtitle{Evidence for a population of microblazars?}
	
	\author{F. Batejat\inst{1}
		\and
		J.~E. Conway\inst{1}
		\and
		A. Rushton\inst{1,2}
		\and
		R. Parra\inst{3}
		\and
		P.~J. Diamond\inst{4,5}
		\and
		C.~J. Lonsdale\inst{6}
		\and
		C.~J. Lonsdale\inst{7}
	}
		
	\institute{Onsala Space Observatory, SE-439 92 Onsala, Sweden; \email{fabien.batejat@chalmers.se}
		\and
		European Southern Observatory, Karl-Schwarzschild-Str 2, 85748 Garching, Germany
		\and
		European Southern Observatory, Alonso de Cordova 3107, Casilla 19001, Santiago 19, Chile
		\and
		CSIRO Astronomy and Space Science, PO Box 76, Epping, NSW 1710, Australia
		\and
		JBCA, School of Physics \& Astronomy, Manchester University, M13 9PL, UK
		\and
		MIT Haystack Observatory, Westford MA, USA
		\and
		North American ALMA Science Center, NRAO, Charlottesville, VA, USA
	}
	
	\date{Received 16 mars 2012 / Accepted 10 mai 2012}
	
	\abstract
		{The two nuclei of the starburst galaxy Arp220 contain multiple compact radio sources previously
		identified as radio supernovae or supernova remnants.}
		{In order to search for an embedded radio AGN, or other possible exotic objects, we have carried out
		a program of VLBI monitoring at 6~cm over three epochs each separated by four months.}
		{Combining the new data with existing data at 6~cm and 18~cm (spanning 4 and 12~years
		respectively) we are able to characterise source flux density variability on a range of time-scales.
		Additionally we analyse the variability of sources in shape and position.}
		{We detect rapid ($< 4~\mathrm{months}$) variability in three sources (W7, W26 and W29).
		These sources show possible superluminal motion ($> 4c$) of jet-like features near
		rapidly varying almost stationary components. These enigmatic sources might be associated with an AGN or a
		highly beamed microquasar (i.e. microblazar). Other hypotheses include that the apparent
		variability is intrinsic and is produced by neutron star powered central components within a supernova remnant, by a
		sequence of several supernovae within super star clusters, or is extrinsic and
		is produced by Galactic interstellar scintillation of very compact non-varying objects.}
		{A microquasar/microblazar origin seems to be the best explanation for the nature of the variable sources
		in Arp220.}
		
	\keywords{Galaxies: nuclei - Galaxies: starburst - Galaxies: individual: Arp220 - Radio continuum: stars - X-rays: binaries}
	
	\maketitle
	
	\section{Introduction} \label{se:introduction}
\object{Arp220} is the result of the merger of two galaxies \citep{NORRIS88} and has two distinct nuclei separated in projection by $\sim 364~\mathrm{pc}$ \citep{SCOVILLE98}. It is the closest ULIRG (Ultra Luminous Infra-Red Galaxy) located 77~Mpc away with a far infrared luminosity \mbox{$\mathrm{L_{FIR}} \sim 1.3\times10^{12}\ \mathrm{L}_{\odot}$} \citep{SOIFER87}, more typical of star-forming galaxies at redshift $z = 1$. It has a similar star-formation density per unit area as redshift $z = 6$ proto-galaxies \citep{WALTER09}.
\citet{SMITH98} made the first detection at 18~cm of compact radio objects associated with the starburst. These objects were detected at shorter wavelengths (13, 6 and 3.6~cm) by \citet{PARRA07}. Both this paper and more recently \citet{BATEJAT11} concluded that most compact sources are either radio supernovae (SNe) or supernova remnants (SNRs). However, not all sources were easy to classify. In this paper we concentrate on three variable sources detected in the Western nucleus of Arp220.
In Section~\ref{se:observation} we present the observations and details of the data processing. In Section~\ref{se:results} we describe our results. In Section~\ref{se:discussion} we discuss several hypotheses for the nature of the three sources. Finally, in Section~\ref{se:conclusion} we give our conclusions.

	\section{Observations} \label{se:observation}
		
\begin{figure*} [ht]
\centering
\includegraphics[width=14.9cm]{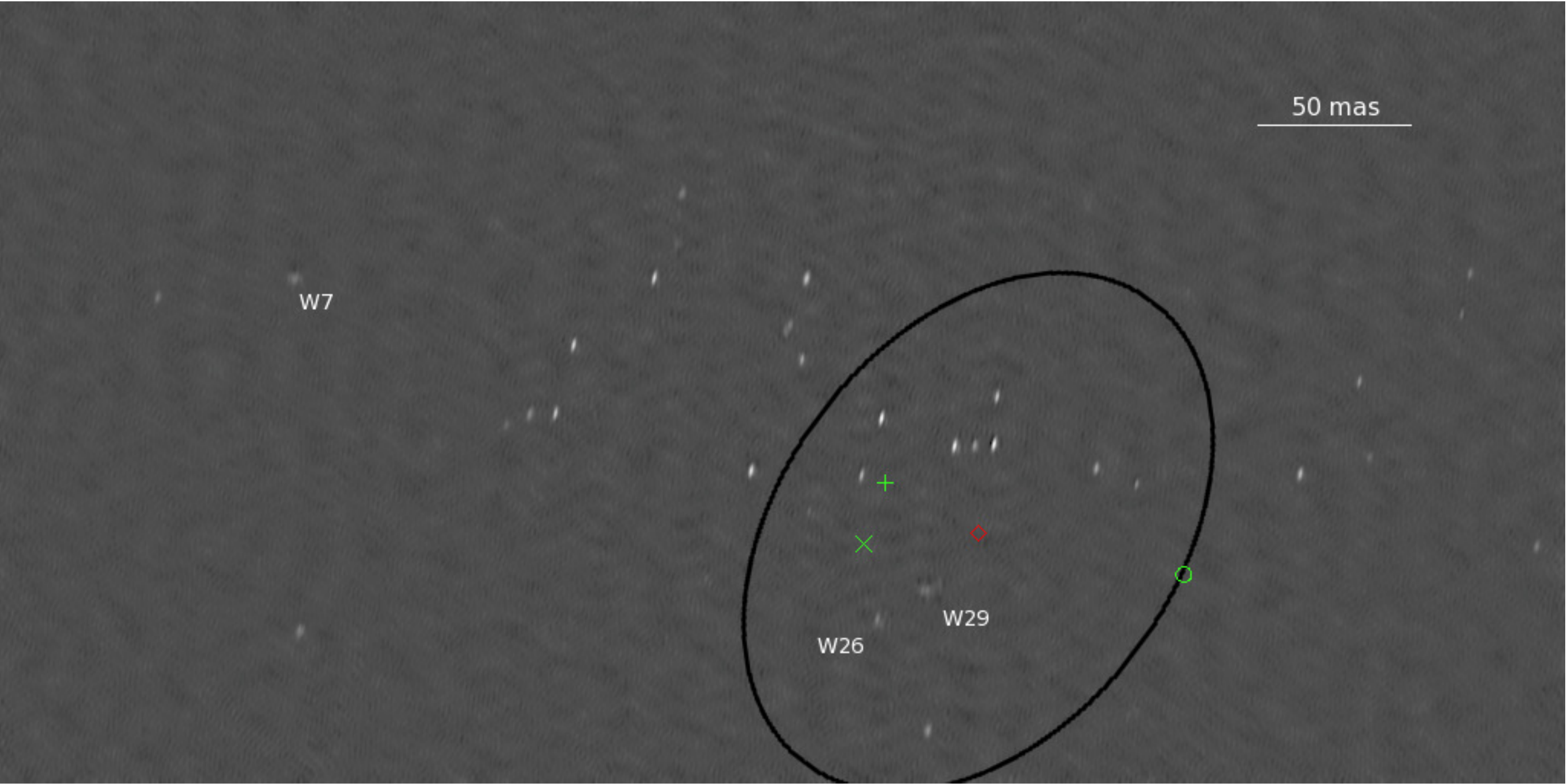}
\caption{Greyscale shows the combined 5~GHz global array GC031A,B,C natural weighted image of the Western nucleus. The noise rms is 8~\uJy/beam, the image size is $512\times256\ \mathrm{mas}$. Labeled in white are the variable sources W7, W26 and W29. The green crosses give the 1.3~mm and 2.6~mm IRAM positions of the hot dust ``AGN'' feature from \citet{DOWNECK07}, the green circle is the position from the 0.8~mm observation of \citet{SAKAMOTO08}. Errors on these positions are estimated to be of order 50 -- 100~mas. The black ellipse gives the orientation and size of the hot dust feature as fitted by \citet{DOWNECK07} centred at the centroid (red square) of the three millimeter interferometer positions.}
\label{fig:exotic} 
\end{figure*}

We performed deep global VLBI 6~cm observations at three epochs (experiments GC031A, GC031B and GC031C) in June 2008 (epoch~1), October 2008 (epoch~2) and February 2009 (epoch~3) with the goal of looking for short time-scale variability in sources of all kinds, i.e. due to type Ib/c supernovae or AGN candidates. In order to limit bandwidth and time-average smearing effects we chose a time resolution of 2~s with frequency resolution of 1~MHz.
We processed these data in a similar way to \citet{BATEJAT11}.
All epochs have been phase-referenced. The absolute astrometric accuracy is estimated to be $\sim0.5~\mathrm{mas}$ \citep{BATEJAT11}.
By combining the three epochs (each with noise rms $\sim$15~\uJy/beam) we have obtained a very high sensitivity image with a noise rms of about 8~\uJy/beam (Figure~\ref{fig:exotic}).

	\section{Results} \label{se:results}

\begin{figure} [ht]
\centering
\hbox{\hspace{-30pt}
\includegraphics[width=11cm]{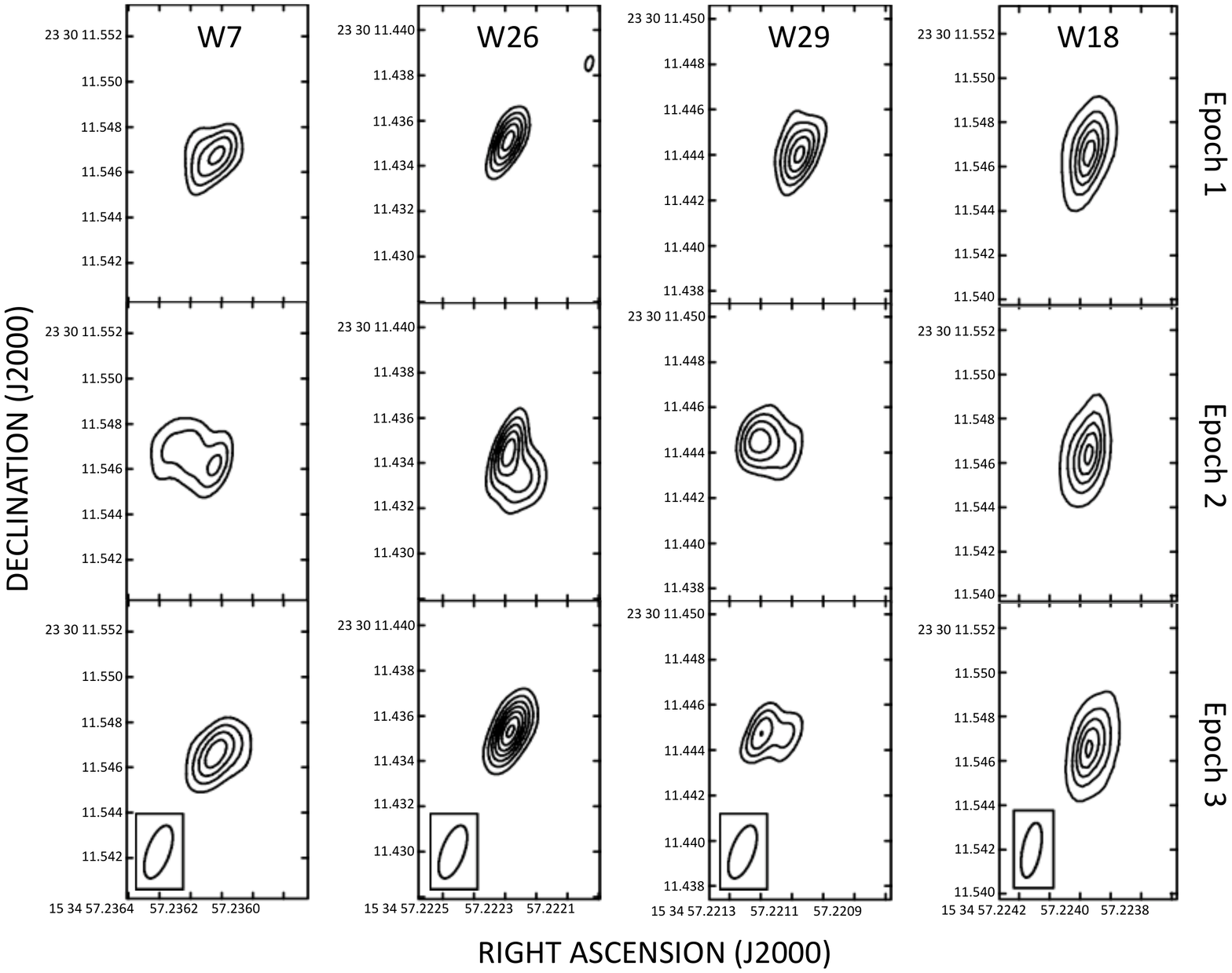}}
\caption{Variability of W7, W26 and W29 over time at 5~GHz. The columns show in order the varying sources W7, W26 and W29 and finally as a control the non-varying resolved source W18 at epochs 1 to 3 (from top to bottom). The contours are at 4, 6, 8, ... times the noise rms for the variable sources and at 4, 12, 20,~... times the noise rms for W18. The noise rms is approximately $15\ \mu\mathrm{Jy/beam}$ at all 3 epochs. The beam is plotted for each source on the last row.}
\label{fig:variable} 
\end{figure}

\begin{figure*} [ht]
\centering
\vbox{\vspace{-5pt}\includegraphics[]{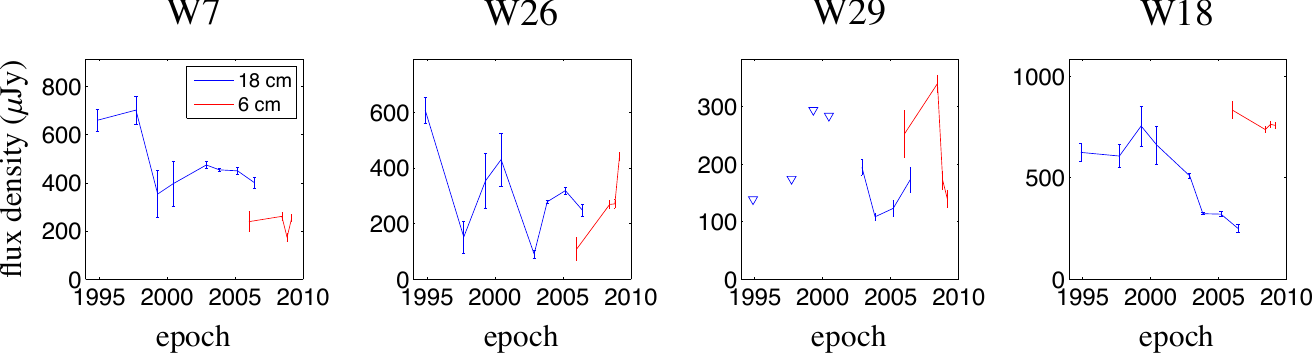}}
\vbox{\vspace{-5pt}\caption{Light curves of W7, W26 and W29 at 18~cm (blue) and 6~cm (red). The SNR candidate W18 is shown as a control source. Error bars are plotted at $\pm 1\sigma$.}}
\label{fig:lc} 
\end{figure*}

Comparison of our three new 6~cm epochs uncovered three rapidly variable sources, W7, W26 and W29 \citep{BATEJAT10}. During 16 years of 18~cm wavelength VLBI monitoring W7 and W26 have always been present \citep[resp. named SN2 and SN9 by][]{ROVILOS05}. W29 has been detected during the last 8 years of high sensitivity observations \citep{LONSDALE06} and the data are consistent with it being present at the same flux density before that. All three sources appear to be persistent, long lived sources. Those sources, all located in the Western nucleus of Arp220, also appear to vary rapidly in position, shape and flux density.
Images of these sources at the three epochs of project GC031 are shown in Figure~\ref{fig:variable} together with W18, a resolved SNR candidate, as a control source.
The light curves of those four sources at 18~cm (1994.87 -- 2006.43) and at 6~cm (2006.02 -- 2009.16) are shown in Figure~\ref{fig:lc}. Gaussian model fitting was applied in the image plane to characterise the source structural variations. In some cases somewhat better fits could be achieved by fitting for more than two Gaussian components or allowing a component to be significantly extended in one direction. In all cases however one or two dominant components were required and their positions and total flux densities were robust independent of the starting model. The parameters of these components are reported below.

W7 contains in all epochs a dominant component with almost constant position. In epoch~1 it had a total flux density of 261~\uJy. 
In epoch~2 this feature's total flux density decreased to 174~\uJy\ 
while in epoch~3 its total flux density increased again to 254~\uJy.
In epoch~2 a secondary feature appeared \mbox{$\sim 1.2~\mathrm{mas}$} away to the East (corresponding to \mbox{$\sim 0.44~\mathrm{pc}$} in Arp220) with total flux density 287~\uJy.
To obtain the best possible fit to the image we require that this secondary component must either be resolved or a third component must exist between the main components. Nothing is detected above the noise of the secondary component at the same position in epochs~1 or 3.

W26 contains in epoch~1 a single dominant component with total flux density 269~\uJy. 
In the subsequent two epochs a bright component persists at the same position with total flux density of respectively 271~\uJy\ and 442~\uJy. 
In epoch~2 a possibly resolved secondary feature briefly appears \mbox{$\sim 1.9~\mathrm{mas}$} to the South-West (corresponding to \mbox{$\sim 0.7~\mathrm{pc}$} in Arp220) with total flux density 218~\uJy. 
Nothing is detected above the noise at the same position in epochs~1 or 3.

W29 contains a single component in epoch~1 with total flux density 340~\uJy.
This component rapidly decreases in flux density in subsequent epochs leaving only a weaker feature with total flux density 172~\uJy\ and 140~\uJy\ respectively, located at approximately the same place. In epoch~2 a much brighter feature lies \mbox{$\sim 1.9~\mathrm{mas}$} to the East (corresponding to \mbox{$\sim 0.7~\mathrm{pc}$} in Arp220) with nothing visible at this position in the earlier epoch. This feature has total flux density 226~\uJy.
In epoch~3 this component maintains its position and flux density (201~\uJy).

	\section{Discussion} \label{se:discussion}

The behaviours of W7 and W26 are similar in the sense that they appear to have a persistent main component varying in flux density with an additional component appearing only at epoch 2. W29 has a single component at epoch 1 which fades away with a new component rising and becoming dominant at subsequent epochs.
If the structural variations are due to components that are ejected these results imply apparent motions with velocities $> 4c$ for W7 and $> 6c$ for W26 and W29. These velocities are similar to the velocity measured in the recently discovered object in M82 \citep{MUXLOW10}. In Arp220 however the sources seem to be persistent while the source in M82 is new. Furthermore, the source luminosity of $\sim 10^{27}~\mathrm{erg~s^{-1}~Hz^{-1}}$ in Arp220 is a factor $\sim100$ times larger than the luminosity of the source in M82.

Alternatively, instead of component motion we could be observing stationary components that independently rise and fade.
Flux variations could be intrinsic but also due to extrinsic Galactic scintillation. At our observation frequency, at the high Galactic latitude of Arp220 ($b=53^{\circ}$) the critical angular scale below which a source shows deep modulation in flux density is $\sim6~\mu\mathrm{arcsec}$ and the predicted timescale is of order 2 hours \citep{WALKER98}.
For $\sim 200$~\uJy\ components scintillation can plausibly contribute to variability. If the component is synchrotron self-absorbed, peaks at 5~GHz and has approximate magnetic field and electron energy equipartition, a source diameter of $\sim 7~\mu\mathrm{arcsec}$ is predicted \citep{CHEVALIER98}. Even smaller source sizes are predicted if the source is relativistically beamed \citep{READHEAD94}.

We note that the hints, found in model-fitting (see Section~\ref{se:results}), that some sources may require at some epochs either more than two compact  components along a line or are extended is an argument against the observed structural variability being explained entirely by scintillation or intrinsic variability of stationary components.  We consider however that the quality of our data is not yet good enough to completely rule out these explanations. Given this, we consider below physical explanations utilising all three variability mechanisms i.e. intrinsic and extrinsic flux variability and emission of jet components.

		\subsection{Supernova remnants and supernovae} \label{se:SNRs&SNe}

Variability within an SNR could be extrinsic and be due to compact scintillating knots (see introduction to Section~\ref{se:discussion}) in an SNR shell. However this would require that a large fraction of the flux density of the SNR is contained in very small knots. Observations of the M82 SNRs \citep{FENECH10} show a much smoother flux density distribution over the whole SNR shell than required to explain the variability in the sources in Arp220.
Alternatively variability within SNRs could be intrinsic and originate from plerion central components powered by neutron stars \citep{BIETBART08}.

Another way to explain intrinsic variability of the sources in Arp220 would be with the rapid evolution of multiple SNe in super star clusters (SSCs). Type Ib/c SNe have radio rise times of $\leq 1~\mathrm{month}$ and can reach the luminosity required to give 100~\uJy\ sources at the distance of Arp220 \citep{CHEVALIER06}. \citet{BONDI12} suggest that source A27 in Arp299 is such a Type~Ib/c SN with rise and decay times $\leq 6~\mathrm{months}$.
Assuming super star clusters with masses $\sim 10^8~\mathrm{M_\odot}$ equal to the most luminous stellar cluster known to date, W3 in NGC~7252 \citep{MARASTON04}, then Starburst99 \citep{LEITHERER95} gives, assuming an instantaneous burst, a maximum {\it total} SN rate of $\sim 0.1~\mathrm{yr^{-1}}$, well below the rate of $\sim 3~\mathrm{yr^{-1}}$ of SN type Ib/c needed to explain our observations. More massive SSCs in Arp220 are ruled out given that the total dynamic mass of the Western nucleus is estimated to be below $10^9~\mathrm{M_\odot}$ \citep{SAKAMOTO99}.

		\subsection{Active Galactic Nuclei} \label{se:AGN}
		
It is possible that at least one of the variable sources is associated with a buried radio-loud AGN within Arp220 similar to that discovered by \citet{PEREZ-TORRES10} in Arp299. The AGN in Arp299 has a luminosity at 5~GHz of $\sim 1.8 \times 10^{27}~\mathrm{erg~s^{-1}~Hz^{-1}}$ comparable to our variable sources. For Arp220 \citet{DOWNECK07} list a number of observations from X-ray to radio consistent with an AGN in the Western nucleus. Additionally they presented dynamical evidence for such a SMBH in the Western nucleus of Arp220 by analysing the position-velocity diagram from CO(2-1) observations. \citet{DOWNECK07} and \citet{SAKAMOTO08} have analysed the mm continuum emission from hot dust and have argued that it may be too compact to originate from a starburst and must instead be black hole powered.
The quoted error bars on the absolute position of this dust feature are large; despite this it is interesting that the mean of these positions (see Figure~\ref{fig:exotic}) lies close to W26 and W29. Those two sources are possible AGN candidates whose apparent structural variability could be due to the combination of superluminal motion in jet components and scintillation of core components.
An obvious problem for the supermassive black-hole (SMBH) interpretation is that we have three variable sources. Furthermore, if beaming is required to explain the variability an even larger parent population is required to explain the presence of three detectable SMBH in Arp220. One way to ease this problem would be to explain variability within W26 and W29 by the scintillation of multiple hotspots in a compact symmetric object (CSO) made up of the two sources. The separation between W26 and W29 is only $\sim7~\mathrm{pc}$. Most CSOs have sizes around 100~pc \citep{AUGUSTO06}. CSOs of the small size of the system W26/W29 would be unusual but not unprecedented.
For completeness we note that the probability that at least one variable source is a background AGN is only $\sim5 \times 10^{-5}$ \citep[estimated based on the detection of 3 compact radio sources of comparable flux densities in the Hubble Deep Field,][]{GARRETT01}.

		\subsection{Microblazars} \label{se:mb}

In this section we consider whether we could explain the variability in the three sources assuming that they are beamed microquasars (i.e. microblazars). Microquasars are accreting stellar mass ($\sim 10~\mathrm{M_{\odot}}$) black holes. An extragalactic microquasar origin has previously been suggested by \citet{MUXLOW10} and \citet{JOSEPH11} for the superluminal source in M82 while \citet{BONDI12} suggests the same origin for a rapidly variable source in Arp299. Assuming core-jet sources with $\Gamma = 5$ observed at an angle to line-of-sight (LOS) of $1/\Gamma$, which is the typical angle for the brightest selected sources in a randomly oriented population, the de-beamed intrinsic (rest-frame) radio luminosity of the Arp220 sources would be $\sim 7 \times 10^{33}~\mathrm{erg~s^{-1}}$ (calculated by integrating up to the observed frequency and assuming a flat spectrum). This is still much larger than the mean hard-state radio luminosity of typical microquasars \citep[$\sim2\times10^{30}~\mathrm{erg~s^{-1}}$,][]{GALLO12} but these have low accretion rates relative to their Eddington accretion rates ($\leq 0.001$). If instead we consider the extremely luminous Galactic microquasar, GRS~1915+105, which accretes at a high fraction of its Eddington rate \citep{RUSHTON10} and consider its most luminous radio state, then the required rest frame luminosity can be reached. Correcting for its observed superluminal motion \citep{FENDER99} and the resulting de-boosting of its jet due to it being orientated close to the sky plane, a rest frame 6~cm luminosity of $\sim 3.4 \times 10^{33}~\mathrm{erg~s^{-1}}$ is estimated. Assuming $\Gamma = 5$ and changing its orientation to an angle $1/\Gamma$ to the LOS, we find that GRS~1915+105 would have an apparent radio luminosity of about half that of the sources in Arp220; for a slightly smaller angle of $0.8/\Gamma$ their luminosities are matched.
Based on this assumption and under the microquasar hypothesis, the double structure observed in the three variable Arp220 sources would be explained as a beamed core and an approaching beamed component.
For a $\Gamma = 5$ source observed at an angle to LOS $1/\Gamma$ the apparent superluminal motion of the latter is of order $5c$, consistent with observations (see introduction to Section~\ref{se:discussion}).

A potential problem with the microblazar hypothesis is the large apparent size of the double sources in Arp220 (0.44 to 0.7~pc) and their long fading times ($< 4~\mathrm{months}$) which are both much larger than for GRS~1915+105 (0.015~pc and tens of days respectively \citep{FENDER99}). For a GRS~1915+105 source seen end-on, the discrepancy in scales/timescales is further exacerbated by the effects of geometric projection and relativistic time compression. 
However, not all microquasar jets are as compact and rapidly varying as in GRS~1915+105. \citet{HAO09} show moving components with $\Gamma = 3$ in the microquasar XTE J1550-564 and other sources extending out to 0.5~pc and persisting for years. If pointed toward us, component decay timescales would be relativistically compressed down to the order of a few months. Linear scales would still be smaller, but of the same order of magnitude, than those observed in the Arp220 varying objects. Another possible explanation for duplicating the source sizes observed in Arp220 would be to have unresolved beamed GRS~1915+105 like objects within clusters; individual sources would vary either intrinsically or because of scintillation due to foreground ionised gas in our Galaxy (see introduction to Section~\ref{se:discussion}).

If we assume that only beamed X-ray binaries with jets having an angle to LOS $\leq 1/\Gamma$ can be detected in the radio, then W7, W26 and W29 represent only 1\% of the parent population of unbeamed X-ray binaries and therefore $\sim 300$ of these are expected in Arp220. The total 2-10~keV soft X-ray luminosity in Arp220 is $L_\mathrm{Arp220} \sim 7 \times 10^{40}~\mathrm{erg~s^{-1}}$ \citep{IWASAWA01,IWASAWA11}  corresponding to $225 \times L_\mathrm{GRS\_1915+105}$ \citep{VERRECCHIA07} and is therefore consistent with this total population size. Finally assuming a star-formation rate of $200~M_{\odot}~\mathrm{yr^{-1}}$ over a continuous burst lasting $> 10~\mathrm{Myr}$ then at near solar metallicity \citet{LINDEN10} predict 2000 X-ray binaries brighter than GRS~1915+105 luminosity, more than enough to explain the required size of the parent population.

	\section{Conclusions} \label{se:conclusion}
		
We have detected three highly variable sources whose structures vary on timescales shorter than 4~months with possible superluminal motion.
It is difficult to conclusively constrain the true nature of those three variable sources.
However, although not all of the properties of the Arp220 variable objects can be matched by any single known Galactic microquasar, their luminosity and possible superluminal motion are similar to an object like GRS~1915+105 seen end-on.
New HSA observations looking for variability on timescales shorter than 30 days are currently being reduced; these will hopefully give us more clues as to the nature of these objects.

	\begin{acknowledgements}

The European VLBI Network is a joint facility of European, Chinese, South African and other radio astronomy institutes funded by their national research councils. The National Radio Astronomy Observatory is a facility of the National Science Foundation operated under cooperative agreement by Associated Universities, Inc. The authors would like to thank the staff at JIVE and NRAO for their diligent correlation of the data.

	\end{acknowledgements}

	\bibliographystyle{aa}
	\bibliography{aa19235-12} 

\end{document}